# Thermal transport properties of the nanocomposite system "porous silicon/ionic liquid"

Pavlo Lishchuk[1], Alina Vashchuk[2*], Sergiy Rogalsky[3], Lesia Chepela[1], Mykola Borovyi[1], David Lacroix[4], and Mykola Isaiev[4]

[1]Faculty of Physics, Taras Shevchenko National University of Kyiv, 64 Volodymyrska street, Kyiv, Ukraine, 01601

[2]E.O. Paton Electric Welding Institute of NAS of Ukraine, 11 Kazymyra Malevycha, Kyiv 03680, Ukraine

[3]V. P. Kukhar Institute of Bioorganic Chemistry and Petrochemistry, National Academy of Sciences of Ukraine, 50, Kharkivske schose, Kyiv 02160, Ukraine

[4]Université de Lorraine, CNRS, LEMTA, Nancy F-54000, France

*Present address: UNIROUEN Normandie, INSA Rouen, CNRS, Groupe de Physique des Materiaux, 76000 Rouen, France

Abstract

This paper investigates thermal transport in a nanocomposite system which is "porous silicon matrix filled with ionic liquid". First, the thermal conductivity and heat capacity of two imidazolium and one ammonium ionic liquids were evaluated using the photoacoustic approach in piezoelectric configuration and differential scanning calorimetry, respectively. Then, the thermal transport properties of the composite system "ionic liquid confined inside porous silicon matrix" were investigated with the photoacoustic approach in gas-microphone configuration. It was found that a significant enhancement of the thermal conductivity of the composite when compared to the individual components, *i.e.* *(i)* more than two times for pristine porous silicon and *(ii)* more than eight times for ionic liquids. These results provide new paths for innovative solutions in the field of thermal management in highly efficient energy storage devices.

1. Introduction

Ionic liquids (ILs) are a class of organic salts presenting melting temperatures below 100 °C. Among all their particular features, ILs show high-voltage stability windows, nearly null volatility, non-flammability, high ionic conductivity, thermal and radiation stability over a wide range of temperatures, corrosion non-activity and recyclability [1]. Moreover, the physicochemical characteristics of ILs can be easily modulated through an endless combination of cationic or anionic constituents, thus leading to labelling them as "designer solvents" [2,3].

Most ILs tend to self-assemble due to competition between electrostatic and Van der Waals interactions of the charged and a-polar alkyl side-chain(s) moieties of their cations [4]. Moreover, the confinement of IL inside a solid porous matrix (host) causes unexpected effects on their physical properties, namely ionic mobility and viscosity [5,6]. In particular, the integration of IL into pores of 20 nm induces a nanometric structuration of the IL molecules, which increases the ionic conductivity by one order of magnitude in comparison to the bulk state [7]. The latter was attributed to changes in ion packing under geometric confinement leading to higher mobility and electrical conductivity. The ILs confinement in unidirectional silica nanopores (7.5–10.4 nm) induces a change of thermal activation behavior from a Vogel-Fulcher-Tamman (VFT) to an Arrhenius-like trend resulting in an enhancement of diffusion coefficients by more than two orders of magnitude [8]. Noteworthy, the effect becomes more pronounced with decreasing pore diameter. Previous works have reported the importance of both the pore's size and the surface chemistry of the host [9].

In designing advanced materials, the confinement of ILs inside porous materials seems to be a promising design strategy for comprehensive application sets. Shared solid porous hosts for ILs are nanoporous polymer [7], nanoporous carbon [10–12], carbon nanotubes [13], silicon [14,15], nanoporous silica [15], silica glass nanocapillaries [16] and metal-organic frameworks [17,18].

Liquids confined in porous matrix systems are promising for their application in various power sources and storage devices [19,20]. In such applications, the systems can be significantly overheated during energy storage and operation. Therefore, understanding thermal properties features in such composite systems is essential for elaborating the new highly efficient energy storage devices.

Porous silicon (PSi) is very attractive as a host for ILs due to its wide diversity of pores [21] and sizeable specific surface area that significantly impact heat and mass transfer of the fluid confined in the porous network. PSi is stiffer than polymer matrix which could lead to significant confinement of the ILs. Moreover, the hydrophobicity of PSi makes it especially attractive as a model for studying the confinement effect of ILs since the water contamination absorbed from the atmosphere is negligible. Although the compatibility of ILs with PSi has been proven [22,23], their thermal transport properties have not yet been investigated. However, their application potential is excellent, as such materials allow the production of parts for cooling circuits, massive thermal isolators, etc. Therefore, the article's primary goal is to investigate the thermal properties of the ILs based nanocomposite systems.

In this work, the hosting matrix is porous silicon fabricated by electrochemical etching. Two imidazolium ILs (hydrophobic/aprotic, hydrophilic/protic) and ammonium IL (hydrophobic/protic) were chosen as fillers. In order to quantify the changes induced by the confinement of ILs on the thermal transport properties of resulting silicon-based composites, the photoacoustic approach was adopted.

2. Materials and methods

Following chemicals were used in this research: imidazole, 1-methylimidazole, lithium bis(trifluoromethylsulfonyl)imide (water solution) (for synthesis), triethylamine hydrochloride (99 %), 1-bromobutane (99 %), bis(2-ethylhexyl) phosphate (97 %) (Sigma-Aldrich), ethyl acetate, methylene chloride (Uoslab, Ukraine).

2.1. Materials fabrication
2.1.1. Synthesis of ionic liquids
2.1.1.1. Synthesis of 1-butyl-3-methylimidazolium bis(trifluoromethylsulfonyl)-imide ([BMIm][TFSI])

Hydrophobic aprotic ionic liquid [BMIm][TFSI] was synthesised according to **Scheme 1**. The mixture of 1-methylimidazole (10 g, 0.12 mol) and 1-bromobutane (25.5 g, 0.13 mol) was stirred at 100 °C for 2 h. After cooling, the viscous liquid was purified by washing with ethyl acetate (3 x 50 ml). Residual solvent was removed in vacuum 15 mbar at 60 °C.

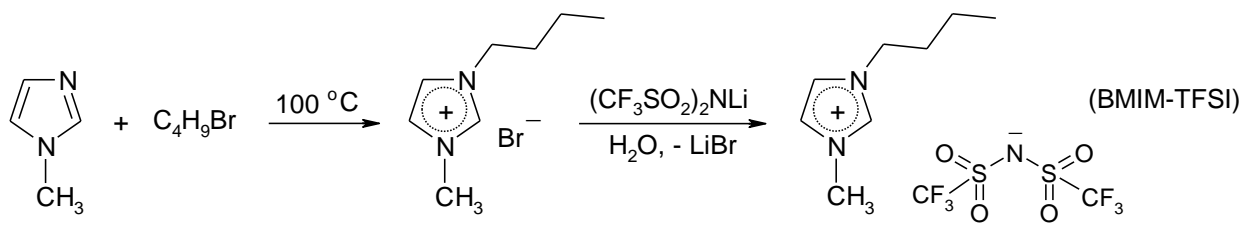

**Scheme 1.** Synthesis of aprotic ionic liquid [BMIm][TFSI]

To the stirred solution of crude 1-butylimidazolium bromide in 100 ml of water was added water solution of lithium bis(trifluoromethylsulfonyl)imide (29 g, 0.1 mol/100 ml). The mixture was stirred for 1 h, and the formed water-immiscible layer was extracted with methylene chloride (2 x 150 ml). The solution was dried with sodium sulfate overnight. Methylene chloride was distilled, the residual solvent was removed in vacuum 5 mbar at 70 °C for 12 h. The liquid product of light brown color was obtained.

$^1$H NMR (400 MHz, DMSO-d$_6$): δ = 0.89 (t, 3H, CH$_3$), 1.24 (m, 2H, NCH$_2$CH$_2$C$H_2$), 1.76 (m, 2H, NCH$_2$C$H_2$), 3.84 (s, 3H, NCH$_3$), 4.15 (t, 2H, NCH$_2$), 7.67 (s, 1H, C$_4$-H), 7.74 (s, 1H, C$_5$-H), 9.06 (s, 1H, C$_2$-H)
$^{19}$F NMR (188 MHz, DMSO-d$_6$): δ = -79.92 (s, 6F, CF$_3$)

2.1.1.2. Synthesis of triethylammonium bis(trifluoromethylsulfonyl)imide ([TEA][TFSI])

Hydrophobic protic ionic liquid [TEA][TFSI] was prepared according to **Scheme 2**. To the stirred water solution of triethylamine hydrochloride (10 g, 0.07 mol/50 ml) was added water solution of lithium bis(trifluoromethylsulfonyl)imide (21 g, 0.07 mol/100 ml). The mixture was stirred for 2 h, and the formed water-immiscible layer was extracted with methylene chloride (2 x 150 ml). The combined organic solution was dried with sodium sulfate. Methylene chloride was distilled, the residual solvent was removed in vacuum 5 mbar at 70 °C. The transparent liquid product was obtained.

$^1$H NMR (400 MHz, CDCl$_3$): δ = 1.31 (t, 9H, CH$_3$), 3.17 (q, 6H, CH$_2$)
$^{19}$F NMR (188 MHz, DMSO-d$_6$): δ = -79.97 (s, 6F, CF$_3$)

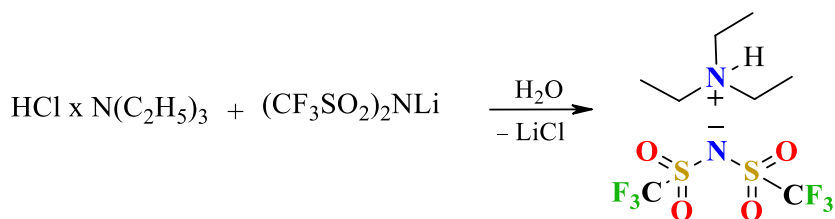

**Scheme 2.** Synthesis of protic ionic liquid [TEA][TFSI]

### 2.1.1.3. Synthesis of imidazolium bis(2-ethylhexyl)phosphate ([Im][BEHP])

Hydrophilic protic ionic liquid [Im][BEHP] was prepared by stirring the mixture of imidazole (5 g, 0.07 mol) and bis(2-ethylhexyl)phosphate (23.5 g, 0.07 mol) at 50 °C for 6 h (**Scheme 3**).

$^1$H NMR (400 MHz, CDCl$_3$): δ = 0.87 (t, 12H, CH$_3$), 1.26-1.44 (m, 16H, CH$_2$), 1.55 (m, 2H, CH$_2$CH), 3.8 (t, 4H, OCH$_2$), 7.11 (s, 2H, C$_4$-H, C$_5$-H), 8.02 (s, 1H, C$_2$-H)

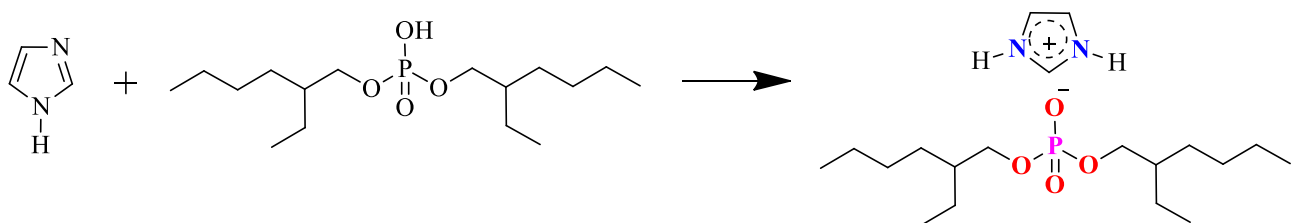

**Scheme 3.** Synthesis of protic ionic liquid [Im][BEHP]

The molecular structure and molecular weight of ILs used in this study are summarised in Table 1.

### 2.1.2. Porous silicon fabrication

As the initial porous matrix for ILs we used mesoporous silicon obtained by electrochemical etching of 500 μm boron-doped p$^+$ type silicon substrate (with a resistivity of 10 - 20 mΩ·cm and [100] orientation). The etching was done in a mixture of hydrofluoric acid (49%) and ethanol in a ratio of 1:1. The porosity (~ 65%) and thickness (50 μm) of PSi layer were controled the etching current density and time and checked by gravimetric method and SEM microscopy, respectivly. The average pore diameter of the obtained samples was found to be about 40 nm.

2.1.3. Fabrication of the nanocomposites

The nanocomposite systems "porous silicon/ionic liquid" were obtained by filling the pores of PSi samples with ILs. This procedure was made by following steps:

1) the IL was added to the PSi surface;

2) the sample was heated to a temperature of 80 °C for 10 min;

3) the samples were passively cooled and kept as it with liquid remaining on the surface for 24 h;

4) excess of IL was removed from the PSi surface.

**Table 1.** Molecular structure and molecular weight of ILs under investigation

| Ionic liquid | Molecular structure | | Molecular weight [g/mol] |
|---|---|---|---|
| | Cation | Anion | |
| *1-butyl-3-methylimidazolium bis(trifluoromethylsulfonyl)imide* [BMIm][TFSI] | 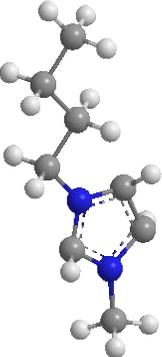 | 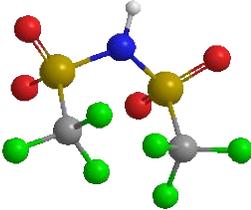 | 419.36 |
| *Triethylammonium bis(trifluoromethylsulfonyl)imide* [TEA][TFSI] | 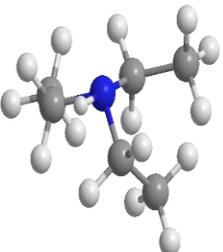 | | 382.3 |
| *Imidazolium bis(2-ethylhexyl)phosphate* [Im][BEHP] | 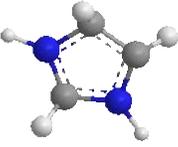 | 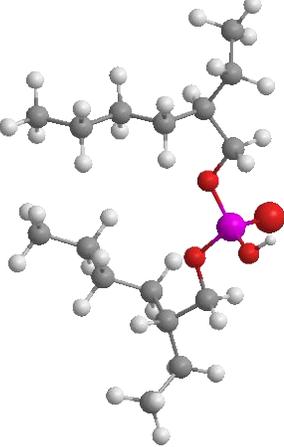 | 390.5 |

For all the samples, the degree of pore filling was determined by the gravimetric method and was not less than 95 %. It was supposed that the PSi matrix was filled completely and homogeneously, the geometry of PSi/IL samples was not modified. [BMIm][TFSI] and [TEA][TFSI] were hydrophobic. Thus, the effect of water contamination absorbed from the atmosphere on thermal conductivities was insignificant.

### 2.2. Experimental setups
#### 2.2.1. Proton nuclear magnetic resonance spectroscopy

Proton Nuclear Magnetic Resonance ($^1$H NMR) spectroscopy was used to confirm the structure of synthesised compounds. The spectra were recorded on a Varian Gemini-2000 (400 MHz) NMR spectrometer.

#### 2.2.2. Density and heat capacity measurements

Information about the structure's density and specific heat capacity plays an essential role in thermal conductivity evaluation by photothermal methods. The densities of all ILs (average value of three measurements) were determined directly from their weighing at room temperature ($T = 25$ °C).

Determination of specific heat of ILs was carried out by the differential calorimetry method. Two thermally insulated aluminium vessels were used so that heat losses to the surroundings can be kept as low as possible. The heat was supplied equally by heating nichrome coils which were immersed simultaneously in vessel with water and with studied liquid. The temperature changes in each vessel was recorded by K-type thermocouples connected to a 2-channel LCD data logger with a resolution of 0.1 °C in a broad range of monitoring applications. A schematic picture of the experimental setup is shown in Fig. 1.

The heat capacity of the studied liquid can be calculated using the known values for Al vessels (calorimetres) and water by the following formula:

$$c_l = \frac{1}{m_l}\left[\frac{(c_{Al}m_{Al}+c_w m_w)(t_2-t_1)}{(t_4-t_3)}\right] - c_{Al}m_{Al} \qquad (1)$$

where $c$ – is a specific heat; $m$ – is a mass; indexes "$Al$", "$w$" "$l$" indicates that the respective parameters are taken for the Al calorimeters (vessels), water in the first calorimeter, and studied liquid in the second calorimeter, respectively; $t_1$ and $t_2$ are the initial and final temperature in the first calorimeter; $t_3$ and $t_4$ are the initial and final temperature in the second calorimeter.

The technique was tested on a control sample PEG 400E, from which the obtained value of the specific heat capacity (2300 J /(kg °C)) correlates qualitatively with the literature data [24]. The results obtained for the studied ILs are shown in Table 2. As can be seen from the table, the specific heat values are within the range of values known from the literature for various ILs [25,26].

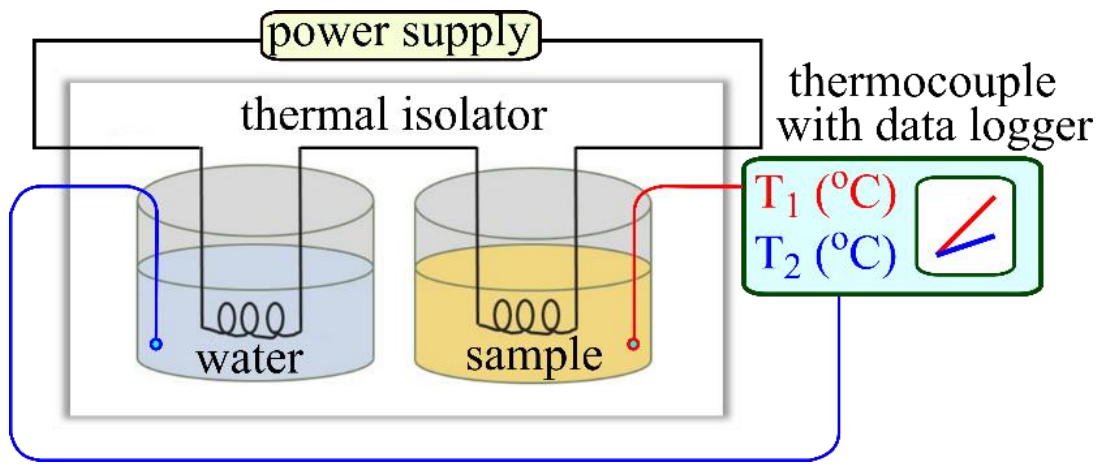

**Figure 1.** Schematic view of the experimental setup for heat capacity measurements of ILs.

**Table 2.** Density and specific heat capacity of ILs

| Liquid | $\rho$ [kg/m³] | $c$ [J /(kg °C)] |
|---|---|---|
| [BMIm][TFSI] | 1380 ± 80 | 1350 ± 110 |
| [TEA][TFSI] | 1380 ± 80 | 1380 ± 110 |
| [Im][BEHP] | 1000 ± 80 | 2060 ± 110 |

2.2.3. Photoacoustic technique for thermal conductivity ($k$) measurements

The photoacoustic technique was used to provide the thermal conductivity of ILs and «PSi/IL» nanocomposite systems [27]. This technique refers to non-destructive methods for investgating the thermophysical properties of materials of different aggregate states and dimensions [28–34]. The experimental setup is described in [35,36]. In our case, installation includes a UV laser (405 nm) that, together with a square signal generator used as a source of modulated light. The UV irradiation is focused on the surface of the samples, creating thermoelastic stresses inside them. Samples inside the PA cell can record the response related to the PA effect. The design and type of PA signal recording depend on the studied object. Namely, the PA cell with a piezoelectric transducer (PZT) was used to study the thermal conductivity of ILs (see Fig. 2(a)), and a gas-microphone photoacoustic cell (PA GM cell) was used for the analysis of composite systems (see Fig. 3(a)). All the experiments by PA technique were done at room temperature.

A) Thermal conductivity measurements of ILs

In this case, the PA cell is a multilayer structure that made of different layers distributed from the laser illumination side in the following order: optically transparent buffer - strongly absorbing layer – liquid sample layer ($l_s$ = 100 μm) – 900 μm aluminium layer – 1000 μm buffer – 700 μm piezoelectric transducer. Thus, thermal energy arises and propagates in the system in the form of thermoelastic stresses due to the absorption of light by a thin absorbing layer located above the liquid. Thermoelastic stresses go through the studied liquid to a metal plate, and finally are recorded by a piezoelectric transducer. The time delay between the start of heating and the system's bending depends on the fluid's thermal conductivity. Thus, to estimate this value, one can use information about the dependence of the voltage on the piezoelectric transducer electrodes on the frequency of optical radiation.

B) Thermal conductivity measurements of nanocomposite systems

To measure the thermal conductivity of composite systems "PSi/IL" we used the classical configuration of PA cell with gas microphone registration [35,36]. This method belongs to the indirect PA research technology, since the signal is recorded in the gas isolated in the PA cell adjacent to the sample surface. Periodic heating of the sample leads to periodic heating of the gas and acoustic waves' appearance, which are recorded by the microphone. Information regarding the frequency response of such a signal indicates the thermophysical properties of the sample under study.

3. Results and discussion
3.1. Thermal properties of ILs

The amplitude-frequency characteristics of the informative signal from the PA cell described in Fig. 2.a were measured in the frequency range from 5 Hz to 18 Hz. The obtained experimental results were analysed by modelling the spatial distribution of variable temperature in the multilayer structure (see Fig. 2.b-d), as described briefly in [31].

The frequency-dependent voltage $U(\omega)$ that occurs at the electrodes of the piezoelectric transducer, in this case, can be written as:

$$U(\omega) \sim \int_{l_b}^{l_{str}} \sigma(z) dz = \int_{l_b}^{l_{str}} \frac{E(z)}{1-v(z)} \int_{l_s}^{l_a} \frac{\alpha_T(s) E \theta(s, \omega)}{1-v(s)} G(z,s) ds dz \qquad (2)$$

where $\omega = 2\pi f$, f is the modulation frequency of the UV light, $\sigma(z)$ is the spatial distribution of the thermoelastic stresses, $\alpha_T(s)$ – is the thermal expansion coefficient of the material, $\theta(s, \omega)$ is the temperature distribution, $E$ – is Young's modulus, $v$ – is Poisson's ratio, $l_{str}$, $l_b$, $l_a$, $l_s$ – are the thicknesses of the piezoelectric transducer, backing material, aluminium and liquid sample layers, respectively.

This equation was used to match the experimental amplitude-frequency characteristics of PA signal when we used some reference liquids as test samples. The thermal conductivity value of fluid was a fitting parameter in the simulation. Before considering ILs, test cases were achieved on water, oil and PEG. The values at which the experiment qualitatively matches the simulation are as follows: for water $K = 0.6 \pm 0.02$ W/(m K), for technical oil $K = 0.15 \pm 0.02$ W/(m K), and for PEG 400E $K = 0.2 \pm 0.02$ W/(m K). These values correlate with technical data of liquids and literature values. This demonstrates the ability of the setup and the inversion technique to recover the thermal conductivity of different liquids. When applied to ILs, the results of PA signal simulation from a multilayer system show that their thermal conductivity is within the range of 0.12 – 0.15 W/(m K). That is typical for ILs [37].

### 3.2. Thermal properties of nanocomposites

PA signal from PSi samples and corresponding "PSi/IL" composite systems were evaluated in the frequency range from 40 Hz to 1000 Hz, in which the PA GM cell is working in the non-resonant mode. The obtained experimental results were analysed by the "critical frequency" method [36]. This method models the thermal perturbations that occur when the sample is irradiated with modulated light by rapidly damped heat waves. They can be characterised by the thermal diffusion length, which depends on the frequency of light modulation and thermophysical properties of the structure:

$$\lambda_T = \sqrt{D_T/(\pi f)} \qquad (3)$$

where $\lambda_T$ is the thermal diffusion length, $D_T$ is the thermal diffusivity of the sample.

In our case, the critical frequency can be defined as "bending frequency" on the amplitude-frequency characteristics of PA signal for a two-layer system, where the upper layer is porous silicon with empty pores or filled with ILs, and the lower layer is monocrystalline silicon. This bending frequency corresponds to the case where the thermal diffusion length coincides with the size of the top layer, dividing the AFC into 2 characteristic frequency regions, according to which the bottom c-Si affects or does not affect the PA response. Thus, the thermal conductivity of the top layer can be obtained from the following expression for its thermal diffusivity:

$$\frac{K}{c\rho} = D_T = \pi f_c l^2 \qquad (4)$$

where $f_c$ – is a critical (bending) frequency.

It should be noted that the volumetric heat capacity of the samples was obtained using the following weighted formula:

$$\begin{aligned} c_{PSi}\rho_{PSi} &= c_{Si}\rho_{Si}(1-\varepsilon) \\ c_{PSi/IL}\rho_{PSi/IL} &= c_{Si}\rho_{Si}(1-\varepsilon) + c_{IL}\rho_{IL}\varepsilon\xi \end{aligned} \qquad (5)$$

where $c_{PSi}\rho_{PSi}$ and $c_{PSi/IL}\rho_{PSi/IL}$ are the volumetric heat capacity of the PSi and "PSi/IL" composite; respectively. $c_{Si}\rho_{Si}$ and $c_{IL}\rho_{IL}$ are the volumetric heat capacity of monocrystalline silicon and ILs, respectively; ε is the porosity, $\xi$ is the degree of filling of the pores with IL.

The experimental amplitude-frequency dependencies of the PA signal for PSi and PSi/IL composites and characteristic critical frequencies are shown on Fig. 3). The estimated averaged thermal conductivity values of the studied ILs, initial PSi samples and the corresponding composite systems "PSi/IL" are shown in Fig. 4.

It was found that the thermal conductivity of the "PSi/IL" composite systems increases up to 2.5 times compared to the initial PSi samples. The difference can be explained by improved thermal contact between the Si crystallites and the IL as a filler [36] and the structuration of the liquid layer close to the interface of the porous matrix [38]. The assessment of these factors to the resulting thermal conductivity was carried out by simulating thermal transport in the original porous silicon and composite systems using COMSOL Multiphysics software. A 3D model reconstructed a porous matrix's morphological features before and after filling with an ionic liquid was created.

As important markers for the simulation of the PSi structure, the average size of silicon nanocrystallites connected by narrow bridges, observed in several SEM images (see Fig. 5A), as well as the porosity of the system, were used. According to the model, a temperature gradient was set in a system with known thermophysical properties of the components (nanostructured silicon with air or ionic liquid) to calculate the generated heat fluxes (see Fig. 5B and 5C). The effective thermal conductivity of the whole structure was evaluated as follows:

$$q = -K_{eff} \cdot \Delta T \qquad (6)$$

where $q$ – is a local heat flux density, $K_{eff}$ – is an effective material's thermal conductivity, $\Delta T$ – is a temperature gradient at the desired direction.

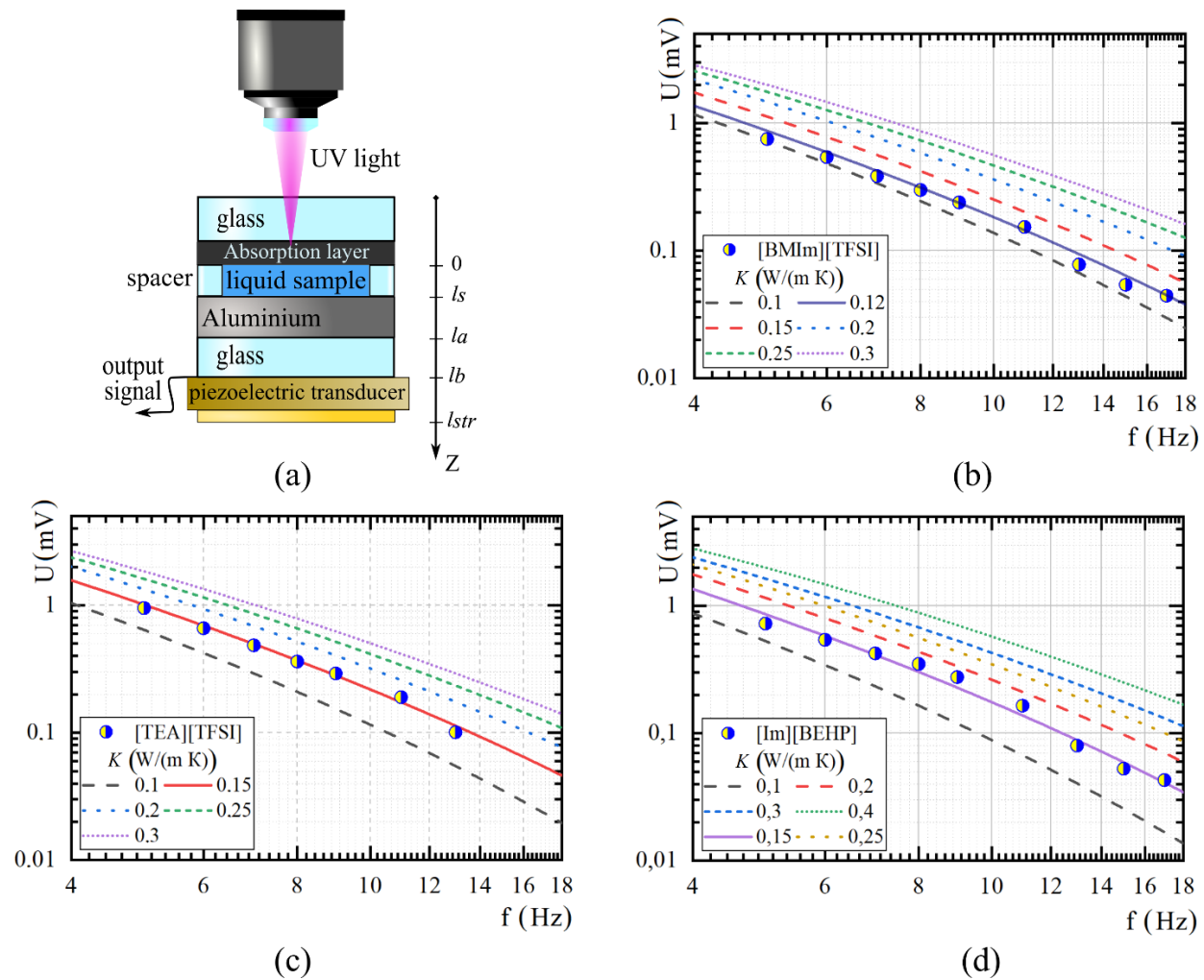

**Figure 2**. PA cell with piezoelectric registration employed (*a*), amplitude-frequency characteristics of a multilayer system with IL: [BMIm][TFSI] (*b*), [TEA][TFSI] (*c*), [Im][BEHP] (*d*). Circles indicate the experimental data, lines – the simulation of PA signal formation at different values of thermal conductivity of the liquid sample.

arXiv:2212.05888https://doi.org/10.48550/arXiv.2212.05888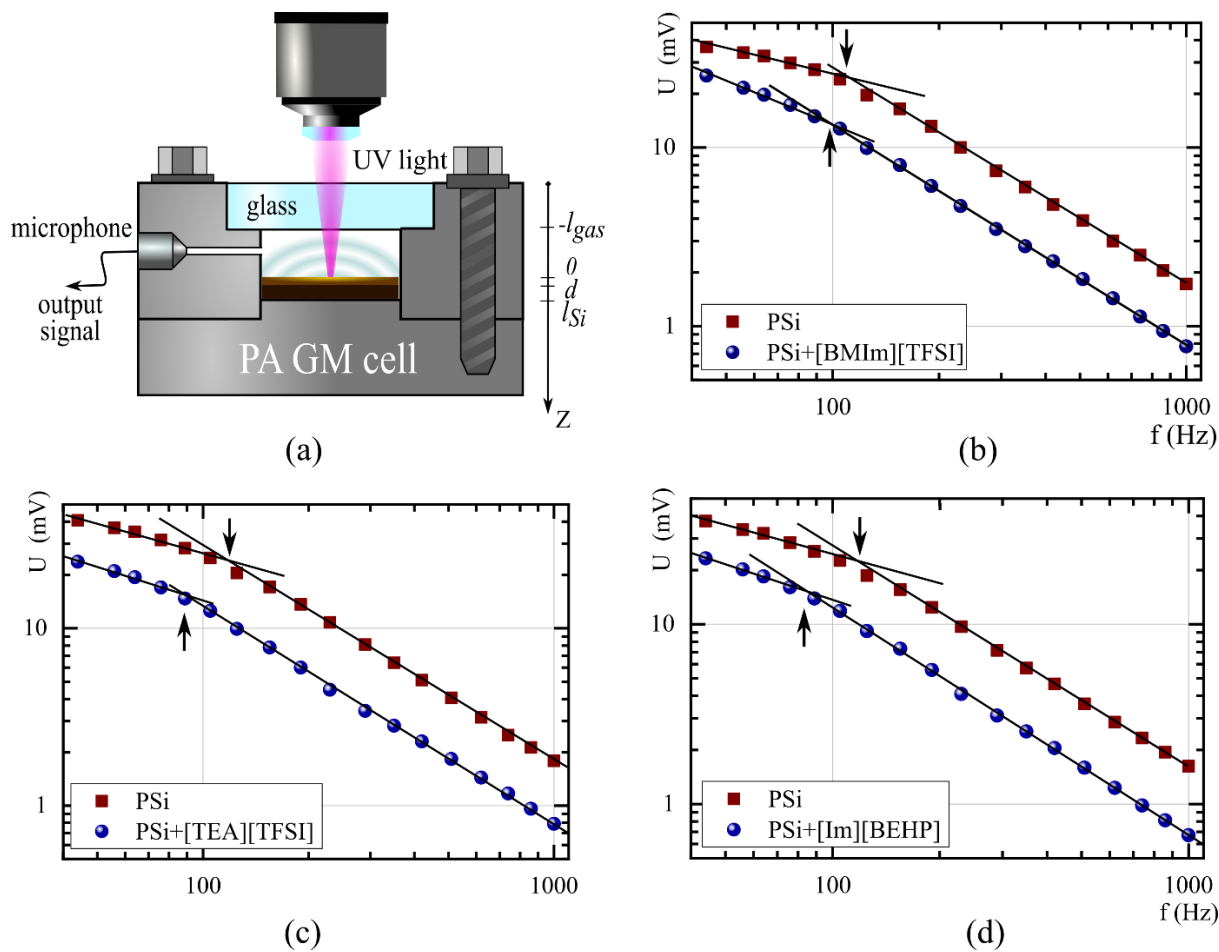

(a)  (b)  (c)  (d)

**Figure 3**. PA cell with gas-microphone registration employed (*a*), amplitude-frequency characteristics of PSi and "PSi/IL" composite systems with IL: [BMIm][TFSI] (*b*), [TEA][TFSI] (*c*), [Im][BEHP] (*d*). Arrows indicate the bending frequency.

The thermal conductivity of Si nanocrystallites as a solid phase in the model material was evaluated according to the Minnich and Chen model, mentioned in [38]:

$$K_{nano-Si} = \frac{K_{c-Si}}{1+SSA\cdot\lambda_x/4} \qquad (7)$$

here $K_{c-Si}$ – is the thermal conductivity of highly doped monocrystalline silicon, $SSA$ – is the specific surface area [39,40], $\lambda_{ph} = 1/\sqrt[3]{N}$ is the linear phonon mean free path, $N$ – is the concentration of boron dopands in silicon structure ( $N$ = (4 – 9.5) ·$10^{18}$ cm$^{-3}$).

The calculated values of the thermal conductivity of nanostructured silicon formed the basis for modelling the effective values of the PSi structure, which turned out to be close in magnitude to the experimentally obtained values within the error. However, in the case of modelling a "PSi/IL" composite system, its thermal conductivity depends on thermal contact resistance between solid/fluid interfaces.

https://doi.org/10.48550/arXiv.2212.05888

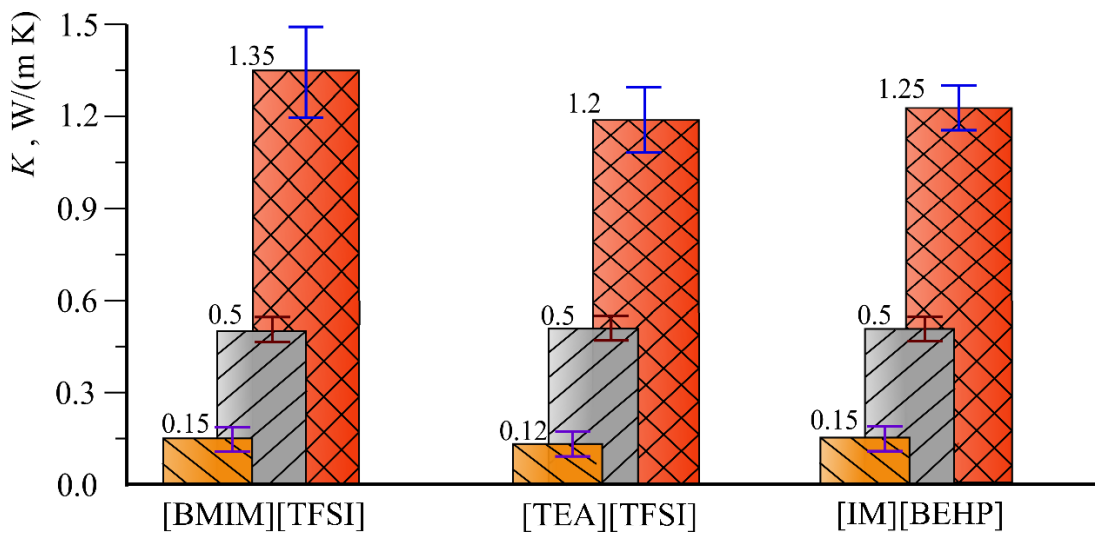

**Figure 4.** Thermal conductivity of IL (\\\-line hatched columns), PSi (///-line hatched columns), "PSi/IL" composite system (#-line hatched columns).

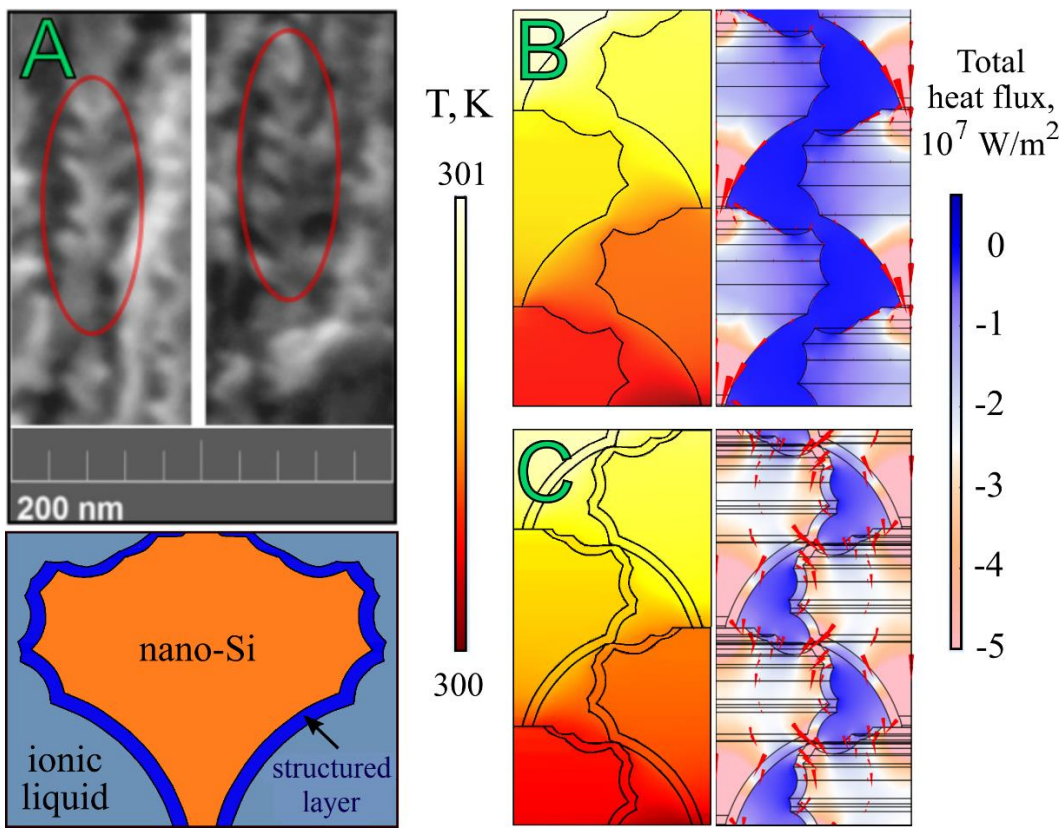

**Figure 5.** SEM cross-sectional images of PSi (highlighted areas show tentative markers indicating the features of the morphology of the porous structure of silicon) and below a schematic view of averaged by size Si nanocrystalline with a structured IL layer around it as the basis for modelling (*a*); schematic view of temperature distribution on the left half and cross-plane total heat flux *q* on the right half in PSi (*b*) and "PSi/IL ([BMIm][TFSI])" composite system (*c*), respectively. Red arrows show the directions of heat fluxes.

In order to match the effective thermal conductivity of the composite with the experimental results, a liquid structuration effect close to the following interface was added in model calculations. As a fitting parameter, we vary the thermal conductivity value of ~1.5 nm thin boundary layer between Si and liquid, where the presence of the surface adsorbed layer of IL with higher density was predicted [41,42].

As a result, the theory gives agreement with experimental data at a thermal conductivity of the boundary layer close to (0.5 – 1.5) W/(m K) that assumes a perfect contact between PSi crystallites and IL with thermal contact resistance $R = (1 – 3)$ m$^2$ K/W.

## 4. Conclusions

This paper investigates the thermal transport properties of the ILs based nanocomposites. In our study, two imidazolium and one ammonium ILs were chosen to combine with the mesoporous silicon fabricated by electrochemical etching of crystalline silicon substrate.

Firstly, we characterised the pristine ILs using differential scanning calorimetry, densimetry and photoacoustic approach with piezoelectric configuration. In such a way, the liquids' density, heat capacity and thermal conductivity were evaluated.

Then, the gas-microphone photoacoustic method was applied to measure the thermal transport properties of PSi/ILs nanocomposite systems. As the main results, the significant enhancement of thermal conductivity of the composite system compared to the pristine matrix and ILs was stated. The structuration of the liquid layer close to the interface of the PSi in composite systems is the most probable reason for this effect. Such structuration leads to modification of thermal transport of the ILs, precisely due to increasing thermal transport along the solid/liquid interface. More specifically, the thermal conductivity of PSi/ILs was ~ 8-10 times higher than bulk ILs and ~ 2.5 times higher compared to PS. Comparing the experimental data and the FEM simulations allow us to estimate the thermal conductance of the structured layer, which corresponds to lower limit of the known literature data regarding simulated interfacial boundary resistance between the ILs and solids. Thus, the results presented in the paper can be used to improve the thermal transport properties of the ILs-based nanocomposite system in various applications connected with energy production, storage, and conversion.

Acknowledgements

MI and DL appreciate the support of EU Stock'NRJ , "Hotline" ANR-19-CE09-0003, and "DropSurf" ANR-20-CE05-0030 projects.